\documentclass[prl,amsmath,amssymb,twocolumn]{revtex4-1}
\usepackage[normalem]{ulem}
\usepackage{amsmath,amssymb}
\usepackage[bookmarks=true,colorlinks,linkcolor=blue,urlcolor=blue,citecolor=blue]{hyperref}
\usepackage{graphicx,amsmath,amsfonts,dsfont}
\usepackage[usenames]{color}
\usepackage{epstopdf}
\usepackage{verbatim}
\usepackage{amsthm}
\usepackage{enumitem}
\hypersetup{colorlinks,linkcolor=blue,filecolor=green,urlcolor=blue,citecolor=blue}

\newcommand{\be}{\begin{equation}}
\newcommand{\ee}{\end{equation}}

\newcommand{\syse}{\end{array}\right.}

\newcommand{\ag}[1]{}

\begin{document}

\title{Driven Markovian Quantum Criticality}

\author{Jamir Marino$^{1,2}$}
\author{Sebastian Diehl$^{1,2}$}

\affiliation{$^1$Institute of Theoretical Physics, TU Dresden, D-01062 Dresden, Germany\\
$^2$Institute of Theoretical Physics, University of Cologne, D-50937 Cologne, Germany}

\begin{abstract}
We identify a new universality class in one-dimensional driven open quantum systems with a dark state. Salient features are the persistence of both the microscopic non-equilibrium conditions as well as the quantum coherence of dynamics close to criticality. This provides a non-equilibrium analogue of quantum criticality, and is sharply distinct from more generic driven  systems, where both effective thermalization as well as asymptotic decoherence ensue, paralleling classical dynamical criticality.  We quantify universality by computing the full set of independent critical exponents within a functional renormalization group approach.

\end{abstract}

\pacs{12.20.Ds, 03.70.+k, 42.50.Ct}

\date{\today}

\maketitle

\emph{Introduction --} There has been a surge of activity in a broad spectrum of experimental platforms, which implement \emph{driven open quantum systems}. In such systems, coherent and driven-dissipative dynamics occur on an equal footing. While such a situation is reminiscent of conventional quantum optics, the systems in point are set apart from more traditional realizations by a large, continuous number of spatial degrees of freedom, giving rise to genuine driven many-body systems. Indeed, experimental realizations range from exciton-polariton systems \cite{Yamamoto10,Carusotto13} over ultracold atoms \cite{Syassen-2008,Carr13,Zhu14}, large systems of trapped ions \cite{blatt12,britton12} and photon Bose-Einstein-Condensates \cite{Weitz10} to microcavity arrays \cite{hartmann08,houck12}. The driven nature at the microscale leads to an intrinsic non-equilibrium (NEQ) situation  even in stationary state due to the explicit breaking of detailed balance. At this microscopic level, the dynamics of such systems is Markovian, i.e. memoryless in time. This is not in fundamental contradiction to genuine quantum effects playing a role, as has been demonstrated theoretically \cite{Diehl08,Verstraete09} and experimentally \cite{Krauter,barreiro-nature-470-486} in many-body systems, where phase coherence or entanglement ensue in the stationary state of tailored driven-dissipative evolution. For the universal critical behavior of such systems, however, despite the fact that  they are ``made of quantum ingredients'', the Markovian character generically leads to a NEQ analogue of classical dynamical criticality: typically, as the result of dissipation, phase transitions in driven-dissipative systems are governed by an emergent effective temperature together with the loss of quantum coherence, and their bulk critical behaviour is captured by equilibrium universality classes \cite{wouters06:_absen, MTKM2006,DallaTorre2010,DallaTorre2012,Oztop2012,DallaTorre2013,Sieberer13,Tauber14, Nagy15, Gor2015}. 

This sparks a natural and fundamental set of questions: Given the intrinsic quantum origin, together with the flexibility in designing such systems, to which extent can effects of quantum mechanical coherence persist asymptotically at the largest distances in the vicinity of a critical point? And if so, what are the precise parallels and differences to criticality in closed equilibrium systems at zero temperature? In other words, is there a driven analogue of quantum critical behavior?

In this work, we address these questions driving a one dimensional open Bose gas with a strong Markovian quantum diffusion, implemented, e.g., with microcavity arrays. When diffusion dominates over the Markovian noise level induced by the environment, a novel critical regime associated to NEQ condensation can be realised, where the coherent quantum mechanical origin of the system and the NEQ driven nature persist at infrared scales.  
%Our model provides a benchmark case in NEQ conditions, for a violation of the mapping among $d$ dimensional critical quantum system and their $d+z$ dimensional counterparts -- a cornerstone of equilibrium phase transitions \cite{sondhi, Sachdev, vojta}, since its critical behaviour cannot be related to the character of a critical semiclassical driven-dissipative Bose gas in three dimensions. 
Our results establish a new driven quantum universality class in one dimension, which we characterize by computing the full set of static and dynamical critical exponents. In particular, we obtain the following key results: (i) \emph{New non-equilibrium fixed point.} The fixed point (FP) associated to quantum NEQ condensation cannot be mapped to the classical FP of driven-dissipative condensation -- as evidenced from novel scaling of the correlation length close to criticality,  and we identify a scaling regime where it governs the fluctuation dominated renormalization group (RG) flow. Due to the fine-tuning of the Markovian noise level in addition to the mass gap, it is less stable - in an RG sense - than a classical fixed point, in analogy to the double fine-tuning of mass and temperature to zero necessary to reach an equilibrium quantum critical point \cite{sondhi}.  (ii) \emph{Absence of decoherence.} In classical equilibrium and driven critical behavior, decoherence causes the fadeout of all coherent couplings in the infrared RG flow, leading to a FP where only purely dissipative dynamics persists. In contrast, the anomalous dispersion relation of the critical modes proper of the novel NEQ fixed point, manifests the simultaneous presence of coherent and diffusive processes. The absence of decoherence is reflected in the degeneracy of the two critical exponents encoding kinetic mechanisms. (iii) \emph{Absence of asymptotic thermalization.} Many driven systems exhibit effective thermal behavior at low frequencies \cite{wouters06:_absen, MTKM2006,DallaTorre2010,DallaTorre2012,Oztop2012,DallaTorre2013,Sieberer13,Tauber14,Gor2015}. The present system does not show this property, and this is characterized by the non-thermal character of the distribution function, as well as  universally by a new independent critical exponent entering the NEQ fluctuation-dissipation relation.  A hallmark of the interplay of these effects is an oscillatory behaviour of the spectral density  as a consequence of the (iv) \emph{RG limit-cycle} behavior of the complex wave function renormalization coefficient.

Finally, notice that the noise in our system is Markovian, in contrast to previous realizations of NEQ quantum criticality \cite{DallaTorre2010, DallaTorre2012}, while the quantum nature of the novel critical regime sharply sets apart our scenario from other NEQ fixed points (FP), as occurring in surface growth \cite{kardar86}, directed percolation \cite{hinrichsen00}, or turbulence \cite{kolmogorov, Kolmo91, berges08,nowak11}.

The results presented here are obtained within a functional renormalization group approach \cite{berges02:_nonper,delamotte08:_introd_nonper_renor_group,litim00:_optim} based on the Keldysh path integral associated to the Lindblad quantum master equation. The nature of the new FP precludes the use of more conventional \emph{classical} dynamical field theories pioneered by Hohenberg and Halperin \cite{HHRev}, and fully necessitates our \emph{quantum} dynamical field theory approach. 

\begin{figure}[t!]\centering
\includegraphics[width=9.0cm] {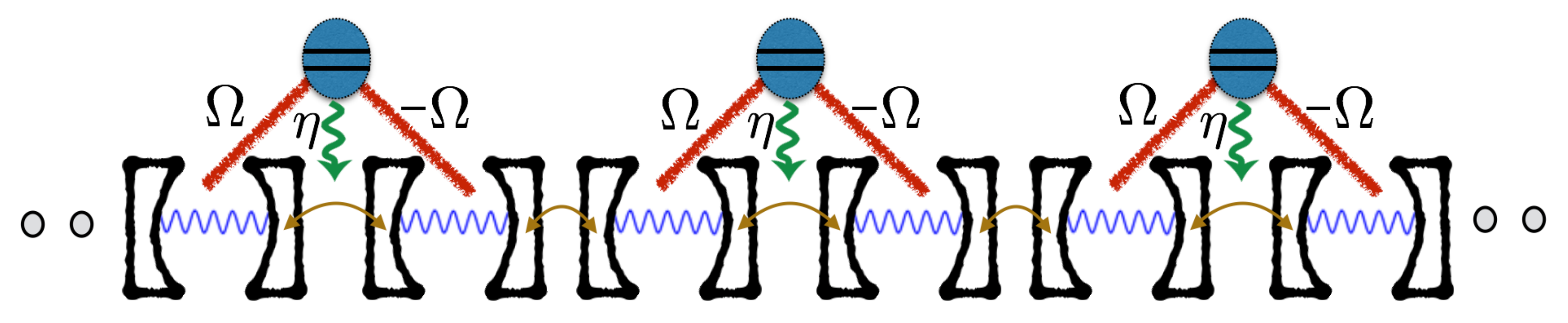}
 \caption{(Color online) A one-dimensional array of microwave resonators, coupled to an array of superconducting qubits (blue dots), which can decay with rate $\eta$ (cavity bosons can tunnel among neighbouring sites - yellow arrows). Each pair of adjacent photonic modes interact with a single qubit via the dipole term, $\mathcal{H}\sim\Omega{\sigma^+_j}(b_i-b_{i+1})+h.c.$;  $b_i$ are the bosonic annihilation operators for the cavity modes and the local qubit Hamiltonians are proportional to the ${\sigma^z_j}$ Pauli matrix. For an energy scale separation $\eta\gg\Omega$, the qubit dynamics can be adiabatically eliminated \cite{Petruccionebook}. This gives rise to Lindblad operators proportional to $\sim b_i-b_{i+1}$, which in the continuum limit yields $L_d(x)$. It imprints an additional diffusion on the propagation of bosons and, crucially for this work, gives rise to a scaling of noise level $\sim q^2$, as  discussed after Eq. \eqref{kinetic}. }
 \label{array}
\end{figure}

\emph{A platform for non-equilibrium quantum criticality --} The starting point for our RG program is the quantum master equation governing the evolution of the density operator, $\hat{\rho}$, of a one-dimensional ($d=1$) bosonic field, $\hat{\phi}(x)$,
\begin{equation}\label{master}
\partial_t\hat{\rho}=-i[H,\hat{\rho}]+\mathcal{L}[\hat{\rho}].
\end{equation}
%Here, coherent quantum many body dynamics, generated by the Hamiltonian $H$, and dissipative processes, encoded in the Liouville super-operator, $\mathcal{L}$, appear simultaneously. 
Two-body collisions of strength $\lambda$, among  bosons (of mass $m$), are encoded in the Hamiltonian, $H$, %is encoded in terms of bosonic degrees of freedom, represented by the field annihilation and  creation operators $\hat{\phi}(x)$, $\hat{\phi}(x)^\dag$, and interacting through ; it reads in one dimension ($d=1$)
%\begin{equation}
%H=\int_x\hat{\phi}^\dag(x)\left(-\frac{\partial_x^2}{2m}\right)\hat{\phi}(x)+\frac{\lambda}{2}\int_{x}\hat{\phi}^\dag(x)^2\hat{\phi}(x)^2.
%\end{equation}
while the Liouvillian can be decomposed into the sum of four dissipative channels $\mathcal{L}=\sum_a \mathcal{L}_a$ ($a=p,l,t,d$), with
$
\mathcal{L}_a[\rho]=\gamma_{a}\int_{x}(\hat{L}_a(x)\hat{\rho} {\hat{L}^\dag_a}(x)-\frac{1}{2} \{\hat{L}^\dag_a(x)\hat{L}_a(x),\hat{\rho}\}),
$
where local Lindblad  operators incoherently create (destroy) single particles $\hat{L}_{p}(x)=\hat{\phi}^\dag(x)$ ($\hat{L}_{l}(x)=\hat{\phi}(x)$), respectively with rates $\gamma_p$ ($\gamma_l$), or destroy two particles, $\hat{L}_{t}(x)=\hat{\phi}(x)^2$, with rate $\gamma_t$. 
The key element of our analysis is the Lindblad operator $\hat{L}_{d}(x)=\partial_x\hat{\phi}(x)$, which is responsible for single particle diffusion with rate $\gamma_d$, and which can be realised through microcavity arrays \cite{hartmann08,houck12} as portrayed in Fig. \ref{array} (see also \cite{Marcos12}).
%, by releasing  the constraint of number particle conservation: Spontaneously decaying superconducting qubits, weakly coupled to the antisymmetric superposition of  microwave  excitations on neighbouring cavities, induce strong diffusion in the quantum master equation for cavity boson excitations. 

In a simple mean field description, when the gain of single particles is balanced by two-body losses, a  condensate $\phi_0=\langle\hat{\phi}(x)\rangle$ with spontaneously chosen global phase can emerge \cite{carusotto05, szymaifmmode06, wouters07}. A crucial ingredient is now the absence of particle number conservation due to pump and loss processes. For this reason, there is no sound mode with dispersion $\omega \sim |q|$ ($z=1$), and, instead, the canonical dynamic exponent is $z=2$. The effective phase space dimension is then $D=d+z=3$, allowing for a condensation transition in one dimension even when fluctuations beyond mean field are taken into account.

A parameter regime of strong quantum diffusion can, indeed, disclose a  quantum critical behaviour analogous to zero temperature quantum criticality, as we are going to glean in the following, recasting the non-unitary quantum evolution encoded in Eq. \eqref{master} into an equivalent Keldysh functional integral formulation of dynamics \cite{kamenevbook,Altland/Simons}. The quadratic part of the action, occurring in the Keldysh partition function, reads 
\be\label{kinetic}
S_{kin}= \int_{t,x} \left( \bar{\phi}_c^{*}, \bar{\phi}_q^{*} \right) \begin{pmatrix}
    0 & \bar{P}^A \\
   \bar{P}^R &
    \bar{P}^K
  \end{pmatrix}
  \begin{pmatrix}
    \bar{\phi}_c \\ \bar{\phi}_q
  \end{pmatrix},
\ee
where $\bar{\phi}_c$ and $\bar{\phi}_q$ are the so-called \emph{classical} and \emph{quantum} fields, defined by the symmetric and antisymmetric combinations of the fields on the forward and backward parts of the Keldysh contour \cite{kamenevbook}. In Eq. \eqref{kinetic}, $\bar{P}^R=(\bar{P}^A)^\dag= i \partial_t + ( {\bar{K}}_R - i \bar{K}_I ) \partial_x^2 + i \bar{\chi}$   is the  retarded (advanced) inverse Green's function, while $\bar{P}^K= i ( \bar{\gamma} - 2{\bar{\gamma}}_d \partial_x^2 )$ is the Keldysh inverse Green's function. In Eq. \eqref{kinetic} we relabelled the parameters in view of RG applications: at the microscopic scale, $k_{UV}$ (the ultra-violet scale   where our RG starts), they are expressed in terms of the couplings entering Eq. \eqref{master}, $\bar{K}_R|_{k_{UV}}\equiv\frac{1}{2m}$, $\bar{K}_I|_{k_{UV}}=\bar{\gamma}_d|_{k_{UV}}\equiv\gamma_d$, $\bar{\chi}|_{k_{UV}}=\frac{\gamma_l-\gamma_p}{2}$ and $\bar{\gamma}|_{k_{UV}}=\gamma_p+\gamma_l$. The existence of two independent Green's functions, $\bar{G}^{R/A}$ and $\bar{G}^K$ -- an exclusive aspect of NEQ dynamics \cite{kamenevbook,tauberbook} -- allows for a distinction between a "retarded mass", $\bar{\chi}$, which controls the distance from the condensation transition, and a "Keldysh mass", $\bar{\gamma}$, which will play in the following the role of a temperature and which microscopically corresponds to a constant Markovian noise level induced by the environment. 

The role of canonical scaling of a  \emph{non-Markovian} quantum noise, $\bar{P}_{eq}^K(\omega)\sim|\omega|$ (responsible for zero-temperature bosonic quantum phase transitions \cite{kamenevbook}), can be taken by the \emph{Markovian} diffusive driving, $\bar{P}^K_{neq}(q)\sim 2\bar{\gamma}_d q^2$, in a model with dynamical critical exponent $z=2$ ($\omega\sim q^z$). In our system, such quantum scaling regime and its associated NEQ fixed point, appear then in the simultaneous limit $\gamma_l\rightarrow\gamma_p$, and $\gamma_l$, $\gamma_p \rightarrow 0$, where  diffusion becomes  dominant over $\bar{\gamma}$  in $P^K$, and the mass gap closes ($\bar{\chi}\to0$). This double fine tuning is analogous to the simultaneous tuning of  spectral gap and  temperature to zero at equilibrium quantum critical points \cite{sondhi}, and it opens the door to the realization of a driven analogue of quantum criticality in our system. In passing we mention that the gapless nature of the NEQ drive  $\bar{P}^K_{neq}(q\rightarrow0)\rightarrow0$, is due to the existence of a \emph{many-body dark state} \cite{Diehl08,Verstraete09}  -- a mode decoupled from noise, located at $q=0$ in our case.

In a strongly diffusive, near critical regime, spectral and Keldysh components of the Gaussian action, Eq. \eqref{kinetic}, scale then alike, $\bar{P}^{R/A/K}(q)\sim q^2$ (in compact notation $[\bar{P}^{R/A/K}]=2$). This fixes the canonical classical and quantum field dimensions to $[\bar{\phi}_c]=[\bar{\phi}_q]=\frac{d}{2}$ and sets the canonical scaling dimension of quartic couplings to $2-d$ (upper critical dimension, $d_c=2$).

We now point out the two key scales which delimit the  scaling regime introduced here and the novel  NEQ quantum critical point discussed later.
The first can be gleaned from an analogy with equilibrium: In the quantum-classical crossover at finite temperature, the quantum scaling associated to equilibrium quantum phase transitions, persists at scales smaller than the de Broglie length, $L_{dB.}\sim\frac{1}{T^{1/z}}$ \cite{sondhi}, where  temperature, $T$, cuts off coherent quantum fluctuations. Analogously, the novel NEQ quantum critical regime is delimited  at low momenta by the Markov momentum scale, $\Lambda_{M}$ -- the threshold where constant Markovian noise prevails over single particle diffusion, spoiling the diffusive scaling of the noise component of the quadratic action ($\bar{P}^K$). For $\Lambda_M$ we find the upper bound $\Lambda_{M}\lesssim 0.2\Lambda_G$ (see Supplemental Material) and at distances larger than $\Lambda_{M}^{-1}$, critical properties are governed by a FP in the Kardar-Parisi-Zhang universality class \cite{AltmanPRX}. 
The second key scale is the Ginzburg momentum scale, $\Lambda_G\simeq\gamma_t/\gamma_d$: at momenta lower than this scale corrections to canonical scaling become effective, indicating the breakdown of a mean-field description \cite{Amit/Martin-Mayor}. According to this analysis, the novel critical behaviour manifests then in the momenta window, $\Lambda_M\lesssim q\lesssim\Lambda_G$.

\emph{Non-Equilibrium Functional Renormalization--}  We now aim at determining the universality class, i.e. the full set of critical exponents associated to the quantum NEQ critical regime,
which is technically characterized  by the so-called Wilson-Fisher FP of the RG equations \cite{Amit/Martin-Mayor}. To this end, we dress the microscopic coefficients of Eq. \eqref{master} with RG corrections, employing a functional RG (FRG) suited for open NEQ quantum many body systems \cite{Sieberer13} (and previously  developed for NEQ closed settings \cite{Jakobs2007,gasenzer08:_towar,berges09:_nonth}). FRG allows us to interpolate from the microscopic dissipative  action to the infrared effective action, introducing an infrared regulator, $\bar{R}_k$, which suppresses stepwise fluctuations with momenta less than an infrared cutoff scale $k$. In this way we can approach smoothly the critical point where infrared divergences govern the physics. The FRG flow is based on a functional differential equation \cite{berges02:_nonper} for the effective action $\Gamma_{k}$, 
%\begin{equation}\label{Wetterich}
$\partial_k\Gamma_k=\frac{i}{2} \operatorname{Tr} \Big[(\Gamma_k^{(\bar{2})}+\bar{R}_k)^{-1}\partial_k\bar{R}_k\Big],$
%\end{equation}
where the trace operation, $ \operatorname{Tr}$, denotes summation over internal degrees of freedom as well as summation over frequencies and momenta, and $\Gamma_k^{(\bar{2})}$ the second functional derivative of the effective action with respect to the fields.
In order to convert the functional differential equation for $\Gamma_{k}$ into a closed set of non-linear differential equations for the RG running of the couplings (the beta functions \cite{Amit/Martin-Mayor, berges02:_nonper}), we provide a functional ansatz for $\Gamma_k\equiv S_{Q,k}=S_{kin}+S_{int}$, where we systematically take into account in $S_{int}\equiv S_{h}+S_{a}$ all  operators which are classified relevant according to the quantum power counting discussed above:
\begin{equation}\label{int}
\begin{split}
S_{h}&=- \int_{x,t}  \frac{1}{2}\left[\frac{\partial{\mathcal{\bar{U}}}_c}{\partial\bar{\phi}_c}\bar{\phi}_q+ \frac{\partial{\mathcal{\bar{U}}}_c^*}{\partial\bar{\phi}_c^*}\bar{\phi}_q^*+ \frac{\partial{\mathcal{\bar{U}}}_q}{\partial\bar{\phi}_q}\bar{\phi}_c + \frac{\partial{\mathcal{\bar{U}}}_q^*}{\partial\bar{\phi}_q^*}\bar{\phi}_c^*\right],\\
S_{a}&= \int_{x,t}  i\bar{g}_1 (\bar{\phi}_c^{*} \bar{\phi}_c-\frac{\bar{\rho}_0}{2})\bar{\phi}_q^{*} \bar{\phi}_q + i \bar{g}_2 ( \bar{\phi}_q^{*} \bar{\phi}_q)^2+ \\
&\quad\quad\quad- \frac{1}{4} [ \bar{g}_3( \bar{\phi}_c^{*} \bar{\phi}_q )^2 -\bar{g}_3^*( \bar{\phi}_c \bar{\phi}_q^* )^2 ].
\end{split} \end{equation}
 $S_h$ and $S_a$ are   respectively the hermitian and anti-hermitian parts of the interaction action.
The potentials, ${\mathcal{\bar{U}}}_c=\frac{1}{2}\bar{u}_c(\bar{\phi}_c^*\bar{\phi}_c-\bar{\rho}_0)^2$ and ${\mathcal{\bar{U}}}_q=\frac{1}{2}\bar{u}_q(\bar{\phi}_q^*\bar{\phi}_q)^2$, have associated complex couplings    $\bar{u}_{c,q}\equiv\bar{\lambda}_{c,q} + i \bar{\kappa}_{c,q}$, which microscopically  coincide with the parameters entering the master equation ($\bar{\lambda}_c|_{k_{UV}}=\bar{\lambda}_q|_{k_{UV}}=\lambda$, $\bar{\kappa}_c|_{k_{UV}}=\bar{\kappa}_q|_{k_{UV}}=\gamma_t$, ${\bar{g}_1}|_{k_{UV}}=2\gamma_t$).
The couplings $\bar{g}_2$  and $\bar{g}_3\equiv\bar{\lambda}_{3}+i\bar{\kappa}_{3}$ are, instead, only generated in the course of renormalization.
In Eq. \eqref{int} we introduced the condensate density (resulting from balance of particles gain/losses), $\bar{\rho}_0$, since in our practical calculations we  approach the transition from the ordered phase, taking the limit of the stationary state condensate $\bar{\rho}_0=\bar{\phi}_c^*\bar{\phi}_c|_{ss}=\bar{\phi}_0^*\bar{\phi}_0\rightarrow0$. In this way, we capture two-loop effects \cite{berges02:_nonper} necessary to compute the full set of critical exponents and thus determine the universality class. We also rewrite the inverse R/A propagators allowing for a complex wave-function renormalization coefficient $Z$, $\bar{P}^R=iZ^*\partial_t+\bar{K}^*\partial_x^2$, with $\bar{K}\equiv \bar{K}_R+i\bar{K}_I$ and $Z\equiv Z_R+iZ_I$,   \footnote{The retarded mass, $\bar{\chi}$, is effectively generated by the condensate.},  whose anomalous dimension  can acquire  a real and imaginary part, $\eta_Z\equiv\eta_{ZR}+i\eta_{ZI}\equiv-\partial_tZ/Z$. We mark that the \emph{quantum} dynamical field theory, $S_{Q,k}$, has a richer RG operator content than a conventional equilibrium Martin-Siggia-Rose action \cite{kamenevbook, tauberbook} or the semiclassical  model in three dimensions for driven-dissipative condensation \cite{Sieberer13} (where $\bar{\gamma}_d=0$, and $\bar{u}_q=\bar{g}_1=\bar{g}_2=\bar{g}_3=0$ from the outset - on the basis of their RG irrelevance). 

Rescaling all the couplings, $\{g\}$, of $S_{Q,k}$ by the quantum canonical power  counting, we find a FP solution, $\{\tilde{g}^*\}$, of the FRG beta functions in terms of the rescaled variables, $\{\tilde{g}\}$ (see \footnote{See Supplemental Material, which includes Refs. \cite{SiebererLong,MarinoDiehl2016}}). The analysis in the vicinity of the FP gives access to the full set of critical exponents. We will use as a benchmark for the salient physical features of the quantum FP in $d=1$ dimensions ($D=3$), its semiclassical driven Markovian counterpart in $d=3$ dimensions \cite{Sieberer13}. 
%A concise summary of the technical steps, which lead us to disclose the new quantum FP, is provided in the Supplemental Material.

\emph{A Non-Equilibrium Quantum Universality Class.} 

\begin{figure}[t!]\centering
\includegraphics[width=8.3cm] {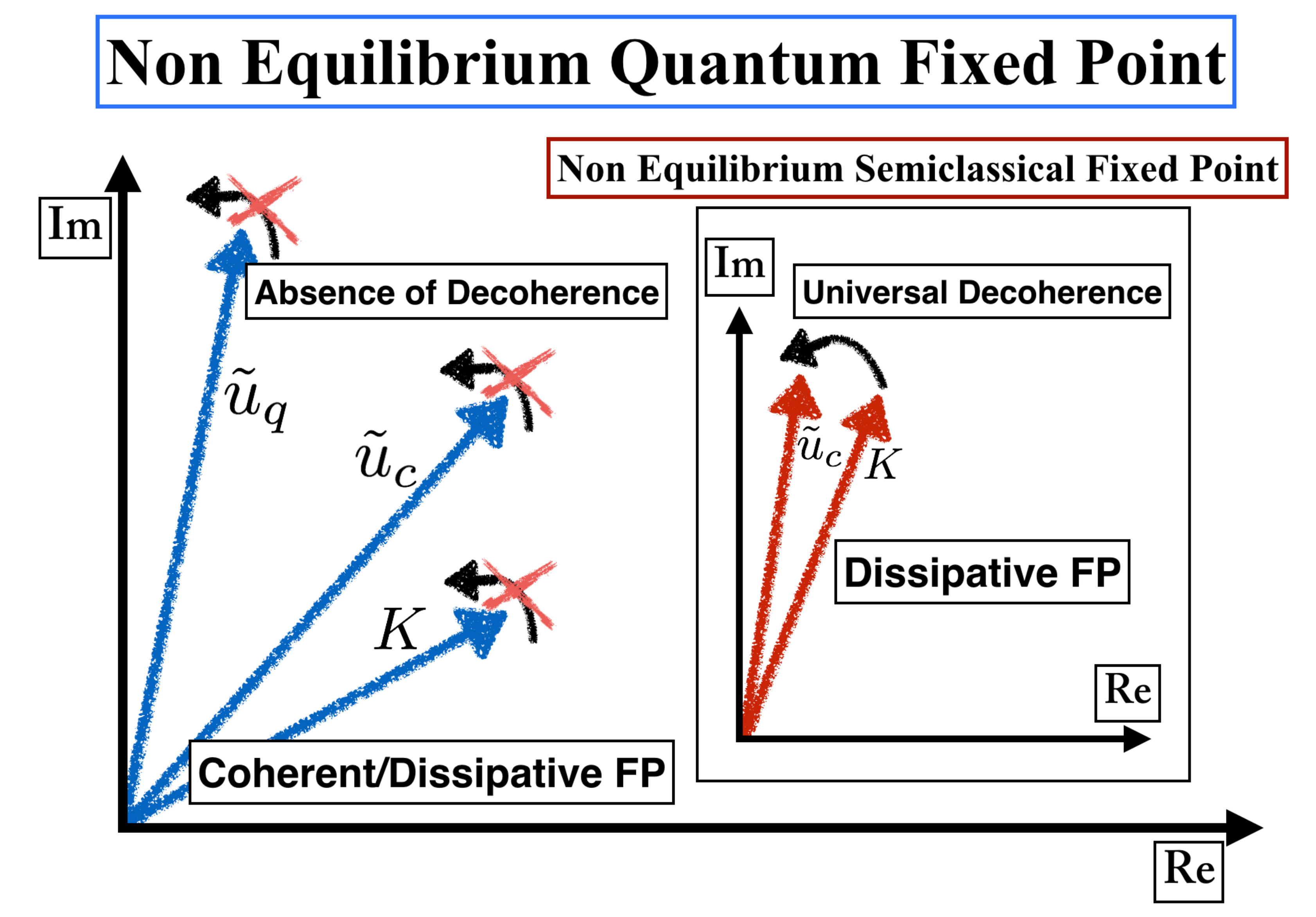}
 \caption{(Color online) In the quantum problem, all the rescaled couplings (we portayed some of the $\{\tilde{g}\}$) keep a non-vanishing real part at the FP and their RG flow freezes in the complex plane (as indicated by the red cross on the curved arrows). In the semi-classical problem, instead, decoherence forces asymptotically all the couplings to flow onto the imaginary axis, and dynamics at infrared scales becomes purely dissipative. }
 \label{Fig2}
\end{figure}

\emph{(i) Non-Equilibrium Fixed Point--} The key new property of the quantum FP is its mixed nature with coexistent coherent and dissipative processes, as shown in Fig. \ref{Fig2}.
This new FP is less stable than the finite temperature FP, or the semi-classical one, since the additional fine tuning of the Markovian noise level is necessary to reach the quantum FP - as discussed above.

In the domain of equilibrium phase transitions, the universality classes of $d$-dimensional critical quantum systems and of their classical $d+z$ dimensional counterparts \cite{sondhi, Sachdev, vojta} coincide. Table \ref{tab} compares the full set of  critical exponents of the quantum transition and of its semiclassical driven-dissipative counterpart \cite{Sieberer13}, elucidating that the analogy does not hold in the case of our NEQ setting. In the vicinity of the transition, the exponent ($\nu$) controlling the divergence of the correlation length of the Bose field, exhibits, for instance, the mismatch among the two critical characters. 

\begin{table}[]
\centering
\begin{tabular}{|l|l|l|l|l|l|l|l|l|}
\hline
Crit. Exps. &$\nu$ & $\eta_{K_R}$ & $ \eta_{K_I} $ &  $\eta_{ZR} $  & $\eta_{ZI} $  &  $\eta_{\gamma_d}$ & $\eta_{\gamma}$   \\ \hline
Quantum &0.405  &-0.025  &  -0.025&0.08 & 0.04 &  -0.26 & $\times$    \\ \hline
Semi-classical  &0.72  &-0.22  &-0.12    &  0.16& 0 & $\times$  & -0.16  \\ \hline
\end{tabular}
\caption{Comparison between the critical exponents of the quantum and semi-classical driven dissipative models (taken from Ref. \cite{Sieberer13}). In the semi-classical scaling, $\gamma\sim k^0$, and the Markovian noise can acquire an anomalous dimension, $\eta_{\gamma}$.} \label{tab}
\end{table}

\emph{(ii) Absence of asymptotic decoherence--} Persistence of quantum mechanical facets at criticality (for length scales shorter than $\Lambda_{M}^{-1}$),  is a common feature  between our FP and equilibrium quantum critical points \cite{sondhi, Sachdev, vojta}. The low energy anomalous dispersion relation of  critical modes, $\omega_k\sim k^{2-\eta_{K_I}}(c_1-ic_2)$, encodes coherent effects ($c_{1,2}$ are two positive constants depending on the quantum FP), in contrast to the purely diffusive leading behaviour of $\omega_k$ in the vicinity of the dissipative FP of the semiclassical model, $\omega_k\sim -ik^{2-\eta_{K_I}}$ \cite{Sieberer13}.
From an RG point of view, the exponent degeneracy, $\eta_{K_R}=\eta_{K_I}=-0.025$, allows for a finite ratio of coherent propagation ($K_R$) versus diffusion ($K_I$), $r=\frac{K_R}{K_I}\sim k^{-\eta_{K_R}+\eta_{K_I}}$, which is thus fully consistent with the results of Fig. \ref{Fig2}, and in particular indicates  absence of decoherence at long distances.

\emph{(iii) Absence of asymptotic thermalization--} The persistence of NEQ character at macroscales and the associated non-thermal character of the distribution function, constitute the 
strongest evidence that the quantum  universality class found in this Letter cannot be related to its  semiclassical driven Markovian counterpart in $d+z$ dimensions, or to an equilibrium FP.

To see this point, we note that the fluctuation-dissipation relation demands, in the 3-$d$ driven-dissipative model, that the effective temperature $T_{C}=|Z|\gamma$ - extracted from the  infrared bosonic distribution function $F_C(\omega,k)\sim\frac{T_{C}}{\omega}$, is scale-invariant. This expresses the principle of detailed balance of thermal equilibrium states (invariance of temperature under the system partition) in an RG language \cite{Sieberer13, DiehlGambx}. Such circumstance  occurs at the semiclassical FP via the emergent exponent degeneracy $\eta_\gamma=-\eta_{ZR}$ (cf. Tab. \ref{tab}) -- the system thermalizes asymptotically.

In the same spirit,  if thermalization were to ensue close to the quantum FP, scale-invariance of the low-frequency distribution function, $F_Q(\omega,k)\sim\frac{T_Q(k)}{\omega}(1+\tilde{\gamma}^*/2)$ ($T_Q(k)\equiv |Z|\gamma_d k^2$), must be expected as a necessary condition. Specifically, replacing the bare scaling of the frequency $\omega\sim k^{z}$ in $F_Q(\omega,k)$, insensitivity to system's partition would manifest in the \emph{exact} scaling relation $F_Q\sim k^0$. The absence of exponent degeneracy, $\eta_{\gamma_d}\neq-\eta_{ZR}$ (cf. Tab. \ref{tab}), signals scaling violation in the infrared behaviour of $F_Q\sim k^{-(\eta_{\gamma_d}+\eta_{ZR})}$, and accordingly the absence of infrared thermalization at the quantum FP.

\emph{(iv) RG limit-cycle of $Z$--} Finally, we notice that the peak of the spectral density -- the imaginary part of the retarded single particle dynamical response, $A(\omega=\operatorname{Re}\omega_k)=\frac{\operatorname{Re}(Z)}{|Z|^2}\frac{1}{\operatorname{Im}\omega_k}$, is sensitive to oscillations present in $ Z\sim k^{-\eta_{ZR}}e^{-i\eta_{ZI}t}$, induced by a non-vanishing $\eta_{ZI}$  ($t=\log (k/\Lambda_{G})$ is the RG flow parameter).
Even if these RG limit-cycle oscillations occur with a huge period, $\frac{2\pi}{\eta_{ZI}}$, they are a remarkable signature of the novel critical behaviour, since they prevent the possibility to have a real wave-function renormalization $Z$, contrary to what happens for purely dissipative relaxational models \cite{HHRev} or for the semiclassical FP, where instead  $\eta_{ZI}=0$ (cf. Tab \ref{tab}) \cite{Sieberer13}.

%\jm{Remarkably, we find a twin FP of the one discussed  above (with $\mathbf{r}^*\rightarrow-\mathbf{r}^*$, and the other rescaled variables unchanged), which exhibits the same critical exponents, except an opposite value of $\eta_{ZI}=-0.04$. It thus displays counter-phase limit-cycle oscillations of $Z$.}

\emph{Conclusions--} We have shown that both quantum mechanical coherence and the microscopic driven nature of open quantum systems can persist close to a critical point, in striking contrast with classical equilibrium and semiclassical NEQ fixed points. We have discussed the impact of this novel critical behaviour on the correlation length of the order parameter, on the distribution function and on the spectral density. An important perspective direction is to study the effect of additional symmetries and conservation laws on the dynamical fine structure of this novel universal behavior.

%Our results pave the way for a systematic classification of universality in driven open systems where genuine quantum effects play a role; for instance, in the spirit of Hohenberg-Halperin models  \cite{HHRev}, we expect additional symmetries and conservation laws to provide a dynamical fine structure in the universal behavior.

\emph{Acknowledgments.} We  thank L. M. Sieberer for support in the early stages of this project, and I. Carusotto for proofreading the manuscript. We also acknowledge useful discussions with F. Becca, M. Buchhold, B. Capogrosso-Sansone, N. Dupuis, T. Gasenzer, A. Kamenev, L. He,  M. Rigol, D. Roscher and M. Vojta. J. M. acknowledges support from the Alexander Von Humboldt foundation. This research was supported by the Austrian Science Fund (FWF) through the START grant Y 581-N16, the German Research Foundation through ZUK 64,  and through the Institutional Strategy of the University of Cologne within the German Excellence Initiative (ZUK 81).

\bibliography{NEQcrit_biblio.bib}

\newpage
\begin{widetext}
\Large\textbf{Supplemental Material of 'Driven Markovian Quantum Criticality'}
\normalsize
\begin{center}
J. Marino$^{1,2}$ and S. Diehl$^{1,2}$

 $^1$ \emph{Institute of Theoretical Physics, TU Dresden, D-01062 Dresden, Germany}

$^2$ \emph{Institute of Theoretical Physics, University of Cologne, D-50937 Cologne, Germany}
\end{center}

\section{Renormalization scheme}
\normalsize
The starting point for the renormalization scheme employed in this Letter is the Wetterich equation [32]
\begin{equation}\label{Wetterich}
\partial_k\Gamma_k=\frac{i}{2} \operatorname{Tr} \Big[(\Gamma_k^{(\bar{2})}+\bar{R}_k)^{-1}\partial_k\bar{R}_k\Big],
\end{equation}
where the functional ansatz for $\Gamma_k$ is the quantum  NEQ Keldysh action, $S_{Q,k}$, of the main text. The trace operation, $ \operatorname{Tr}$, denotes summation over internal degrees of freedom as well as summation over frequencies and momenta; $\Gamma_k^{(\bar{2})}$ denotes the second functional derivative of the effective action  with respect to classical and quantum fields, and finally $\bar{R}_k$ is the optimized cutoff [34], $\bar{R}_k=\bar{K}(q^2-k^2)\theta(k^2-q^2)$, where $\bar{K}$ denotes the bare kinetic coefficient. \\

We now summarize the technical steps which lead us to disclose the Wilson-Fisher FP.  \\

We first introduce renormalized fields $\phi_c=\bar{\phi}_c$, $\phi_q=Z\bar{\phi}_q$;  this choice implies that the complex wave-function renormalization $Z$ that multiplies the time derivative in $\bar{P}^{R/A}$, is absorbed in the field variables, and that kinetic coefficients and couplings are rescaled in the following way: $K=\bar{K}/Z$, $\gamma_d= \bar{\gamma}_d/|Z|^2$,  $u_c= \bar{u}_c/Z$, $u_q= \bar{u}_q/(Z^*|Z|^2)$, $g_1=\bar{g}_1/|Z|^2$, $g_2= \bar{g}_2/|Z|^4$, $g_3= \bar{g}_3/Z^2$, $\gamma=\bar{\gamma}/|Z|^2$.

We project the functional flow encoded in Eq. \eqref{Wetterich}, into the flow equations of the couplings, taking functional derivatives wrt to their linked quartic operators, and evaluating such derivatives on the classical and quantum background fields associated to condensation ($\bar{\phi}_c=\bar{\phi}_c^*=\bar{\phi}_0$ and $\bar{\phi}_q=\bar{\phi}_q^*=0$).

The beta functions expressed in terms of  the RG flow parameter, $t=\log (k/k_{UV})$, read: 
\begin{equation}
  \label{beta}
  \begin{split}    
\partial_t \lambda_c & =\beta_{\lambda_c}=\eta_{ZR} \lambda_c - \eta_{ZI} \kappa_c +\Delta \lambda_c, \\ 
\partial_t \kappa_c & =\beta_{\kappa_c}= \eta_{ZI} \lambda_c + \eta_{ZR} \kappa_c +\Delta \kappa_c, \\ 
\partial_t \lambda_q & =\beta_{\lambda_q}=3\eta_{ZR} \lambda_q +  \eta_{ZI} \kappa_q +\Delta \lambda_q, \\ 
\partial_t \kappa_q & =\beta_{\kappa_q}=3\eta_{ZR} \kappa_q -  \eta_{ZI} \lambda_q +\Delta  \kappa_q, \\ 
\partial_t g_1 & =\beta_{g_1}=2\eta_{ZR} g_1 +\Delta g_1, \\ 
\partial_t g_2 & =\beta_{g_2}=4\eta_{ZR} g_2 +\Delta  g_2, \\ 
\partial_t \lambda_3 & =\beta_{\lambda_3}=2\eta_{ZR} \lambda_3 -  2\eta_{ZI} \kappa_3 +\Delta  \lambda_3, \\ 
\partial_t \kappa_3 & =\beta_{\kappa_3}=2\eta_{ZR} \kappa_3 + 2\eta_{ZI} \lambda_3 +\Delta  \kappa_3, \\ 
\partial_t \gamma & =\beta_{\gamma}=2\eta_{ZR} \gamma +\Delta \gamma, \\ 
\end{split}
\end{equation}
where $\eta_{ZR}$ and $\eta_{ZI}$ are computed along the lines of Ref. [54] and they appear in Eqs. \eqref{beta}  as a consequence of the rescaling of the couplings through the complex wave-function renormalization coefficient, $Z$.  The  computation of $ \Delta\lambda_c$, $...$, $\Delta \gamma$ has been carried extending the scheme of  [54] to extract loop corrections of vertices which are quadratic, cubic, and quartic in the quantum fields. Further technical details of this procedure will be provided elsewhere [55].

The NEQ critical properties under inspection are described by a quantum scaling solution of the flow equations, Eq. \eqref{beta}. We define three dimensionless ratios of coherent to dissipative couplings, $r=K_R/K_I$, $r_U=\lambda_c/\kappa_c$, ${r^Q_U}=\lambda_q/\kappa_q$, and by virtue of the quantum canonical power counting, we introduce the dimensionless retarded $\tilde{m}_R=\frac{2\kappa_c\rho_0}{K_Ik^2}$, and Keldysh mass, $ \tilde{\gamma} = \frac{\gamma}{\gamma_d k^2}$, likewise the three dimensionless effective Keldysh masses $\tilde{m}^1_{\textsl{K}}=\frac{2g_1\rho_0}{\gamma_dk^2}$, $\tilde{m}^2_{\textsl{K}}=\frac{2\lambda_3\rho_0}{\gamma_d k^2}$, $\tilde{m}^3_{\textsl{K}}=\frac{2\kappa_3\rho_0}{\gamma_d k^2}$,  and the dissipative rescaled couplings $
 \tilde{\kappa}_c =\frac{1}{4\pi} \frac{ \gamma_d
    \kappa_c}{K_I^2 k}$, $ \tilde{\kappa}_q =\frac{1}{4\pi} \frac{
    \kappa_q}{ \gamma_d k}$, $ \tilde{g}_2 =\frac{1}{4\pi} \frac{ K_I
    g_2}{ \gamma_d^2k}$.
    The flow equations of these dimensionless rescaled variables acquire a contribution from the running of the anomalous dimensions, $\eta_a=-k\partial_k \ln a$, with $a=K_R,K_I,\gamma_d$, which are functions of the rescaled variables too. 

Employing Eqs. \eqref{beta}, the flow of the ratios, $r_U$, $r_{U}^Q$, can then be written as
\begin{equation}
\label{ratios}
\begin{split}
   \partial_t r_{U} & = \beta_{r_{U}} =
  \frac{1}{\kappa_c} \left( \beta_{\lambda_c} - r_{U} \beta_{\kappa_c} \right),\\ 
  \partial_t {r_{U}^Q} & = \beta_{{r^Q_{U}}} = \frac{1}{\kappa_q}
  ( \beta_{\lambda_q} - {r^Q_{U}} \beta_{\kappa_q} ).\\
  \end{split}
  \end{equation}
  For the ratio of real ($K_R$) and imaginary parts ($K_I$) of the kinetic coefficient, $K$, we have instead
  \begin{equation}\label{erreK}
 \partial_t r  = \beta_{r} =r({\eta}_{K_I}-\eta_{K_R}), 
  \end{equation}
where in the first equation we introduced the anomalous dimensions of the renormalized kinetic coefficient $K$, 
\begin{equation}
\begin{split}
\eta_{K_R}=&\bar{\eta}_{K_R}-\eta_{ZR}+\frac{\eta_{ZI}}{r},\\
\eta_{K_I}=&\bar{\eta}_{K_I}-\eta_{ZR}-\eta_{ZI}r,\\
\end{split}
\end{equation}
which are expressed in terms of $\eta_{ZR}$, $\eta_{ZI}$, and of the anomalous dimensions of the bare kinetic coefficients $\bar{\eta}_{K_R}$, $\bar{\eta}_{K_I}$.  $\bar{\eta}_{K_R}$ and $\bar{\eta}_{K_I}$ are again extracted from the Wetterich equation, Eq. \eqref{Wetterich}, following the procedure discussed in  Appendix C of Ref. [54].\\

For the dimensionless masses of the problem, we have instead 
\begin{equation}
  \label{mass}
\begin{split}
  \partial_t \tilde{m}_R &= \beta_{\tilde{m}_R}=- \left( 2 - \eta_{K_I} \right) \tilde{m}_R + \frac{\tilde{m}_R}{\kappa_c}
  \beta_{\kappa_c} + \frac{2 \kappa_c}{K_I k^2 } \beta_{\rho_0},\\
 \partial_t {\tilde{m}^1_K} &= \beta_{\tilde{m}^1_K}=- \left( 2 -\eta_{\gamma_d} \right) \tilde{m}^1_K +\frac{2}{\gamma_d k^2}(\rho_0\beta_{g_1}+g_1\beta_{\rho_0}),\\
  \partial_t {\tilde{m}^2_K} &=\beta_{\tilde{m}^2_K}=- \left( 2 -\eta_{\gamma_d} \right) \tilde{m}^2_K +\frac{2}{\gamma_d k^2}(\rho_0\beta_{\lambda_3}+\lambda_3\beta_{\rho_0}),\\
 \partial_t {\tilde{m}^3_K} &=\beta_{\tilde{m}^3_K}=- \left( 2 -\eta_{\gamma_d} \right) \tilde{m}^3_K +\frac{2}{\gamma_d k^2}(\rho_0\beta_{\kappa_3}+\kappa_3\beta_{\rho_0}),\\
  \partial_t \tilde{\gamma} &=\beta_{\tilde{\gamma}}=- \left( 2 -\eta_{\gamma_d} \right) \tilde{\gamma} +\frac{1}{\gamma_d k^2}\beta_\gamma.
\end{split}
\end{equation}
where  loop corrections to the condensate amplitude make their appearance in $\beta_{\rho_0}$ (computed following Ref. [54]), as well the anomalous dimension for the diffusion coupling in the Keldysh sector
\begin{gather}
\eta_{\gamma_d}={\bar{\eta}}_{\gamma_d}-2\eta_{ZR}.
\end{gather}
${\bar{\eta}}_{\gamma_d}$ is determined by loop corrections to $\bar{\gamma}_d$, and it is computed generalizing the technique developed in [54] for the extraction of the bare kinetic coefficients $\bar{\eta}_{K_R}$, $\bar{\eta}_{K_I}$ [55].\\

Finally, we need the flow equations of the following dimensionless dissipative couplings 
\begin{equation}\label{coup}
\begin{split}
  \partial_t \tilde{\kappa}_c &= \beta_{\tilde{\kappa}_c} = - \left( 1 - 2 \eta_{K_I}
    + \eta_{\gamma_d} \right) \tilde{\kappa}_c + \frac{1}{4\pi}\frac{  \gamma_d}{  {K^2_I}k}
  \beta_{\kappa_c}, \\
   \partial_t \tilde{\kappa}_q&= \beta_{\tilde{\kappa}_q}= -\left( 1- \eta_{\gamma_d} \right) \tilde{\kappa}_q +  \frac{1}{4\pi}\frac{1}{\gamma_d k} \beta_{\kappa_q},\\
   \partial_t \tilde{g}_2 &= \beta_{\tilde{g}_2}=- \left( 1 +  \eta_{K_I}
    -2 \eta_{\gamma_d} \right) \tilde{g}_2 + \frac{1}{4\pi} \frac{ K_I }{ \gamma_d^2k}
  \beta_{g_2}.
  \end{split}
\end{equation}
Setting to zero the RHS of \eqref{ratios},\eqref{erreK},\eqref{mass},\eqref{coup}, we find the quantum FP  with coexistent coherent and dissipative processes: $r^*=1.55, r^*_U=0.397, {r^Q_U}^*=-0.248, \tilde{\kappa}_c^*=0.146, \tilde{\kappa}_q^*=-0.028, \tilde{g}^*_2=0.0047,   \tilde{m}^*_R=0.900, \tilde{\gamma}^*=0.023, {\tilde{m}^{1*}}_{\textsl{K}}=0.22, {\tilde{m}^{2*}}_{\textsl{K}}=-0.0484, {\tilde{m}^{3*}}_{\textsl{K}}= 0.803$. 

Linearization around the fixed point provides the stability matrix, containing the information on the critical exponents. The number of non-degenerate independent blocks in this matrix gives the number of independent critical exponents,  which are five in our case (cf. Tab. 1 in the main text).

\section{Estimate of $\Lambda_M$}

At equilibrium, $\Lambda_{dB.}$ is determined from the outset, since  temperature is RG invariant. 
In our NEQ problem, the  Markovian noise level $\gamma(k)$ is, instead, scale dependent and its value can be renormalized by loop corrections.

After the RG flow enters the fluctuation dominated regime, $k\lesssim\Lambda_G$, loop corrections can sensitively affect the value of $\Lambda_{M}$. The aim of the second part of our estimate is to extract the functional dependence of $\gamma(k)$ in  proximity of the departure from the  FP. In order to accomplish this goal, we consider the one-loop beta function for the Markovian noise level, where we rescale all the couplings besides $\gamma$ itself in view of replacing them by their FP values,
\begin{equation}\label{flowgamma}
k\partial_k \gamma=-4\tilde{g}_1^*\frac{3\gamma+2k^2\gamma_d}{3(1+\tilde{\chi}^*)^2},
\end{equation}
where $\tilde{g}_1=\frac{1}{4\pi}\frac{g_1}{K_Ik}$ and $\tilde{\chi}=\frac{\chi}{K_Ik^2}$.
The solution of Eq. \eqref{flowgamma}, $\gamma(k)=-b\gamma_dk^2+Ck^{-a}$, contains the FP-dependent parameter, $a=4\frac{{\tilde{\kappa}_c}^*{\tilde{m}^{1*}}_{\textsl{K}}}{\tilde{m}^*_R(1-\tilde{m}^*_R/4)^2}$ (with $b=\frac{2a}{3(2+a)}$), and the integration constant $C=\left(b+\tilde{\gamma}^*\right)\gamma_d\Lambda_G^{2+a}$, which  has been evaluated imposing the boundary condition for the differential equation Eq. \eqref{flowgamma} around the Ginzburg scale,  $\gamma(k\sim\Lambda_G)=\tilde{\gamma}^*\gamma_d\Lambda_G^2$.

Running couplings acquire rapidly their fixed point values when the flow enters the fluctuation-dominated regime, $k\lesssim\Lambda_G$. This justifies both the replacement by their FP values in Eq. \eqref{flowgamma}, and the computation of the integration constant $C$. 

Solving $\gamma(k)\simeq2\gamma_dk^2$, we find then
\begin{equation}\label{markov}
k\simeq\Lambda_G\left(\frac{{\tilde{\gamma}}^*+b}{2+b}\right)^{\frac{1}{2+a}}.
\end{equation}

Replacing the FP values of the rescaled variables in Eq. \eqref{markov}, we find $\Lambda_{M}\lesssim0.25\Lambda_G$.

We also performed a more detailed numerical estimate of  $\Lambda_M$ using the above strategy but based on the two-loop beta functions. The result does not change significantly, $\Lambda_{M}\lesssim0.2\Lambda_G$, which is the value reported in the main text.  For this reason, we presented here the technically less involved and physically more transparent  estimate of $\Lambda_M$ from the one-loop beta function of $\gamma$.\\
\end{widetext}

\end{document}